\documentclass[prl,twocolumn,nofootinbib]{revtex4}
\usepackage{graphicx}
\usepackage{latexsym}
\def\be{\begin{equation}}
\def\ee{\end{equation}}
\def\bea{\begin{eqnarray}}
\def\eea{\end{eqnarray}}

\begin{document}
\title{Weighing neutrinos in the presence of a running primordial spectral index}
\author{Bo Feng${}^{a}$}

\author{Jun-Qing Xia${}^{b}$}

\author{Jun'ichi Yokoyama${}^{a}$}

\author{Xinmin Zhang${}^{b}$}

\author{Gong-Bo Zhao${}^{b}$}

\affiliation{ ${}^a$ Research Center for the Early Universe(RESCEU),
Graduate School of Science, The University of Tokyo, Tokyo 113-0033,
Japan}

\affiliation{${}^b$Institute of High Energy Physics, Chinese Academy
of Science, P.O. Box 918-4, Beijing 100049, P. R. China}

\begin{abstract}

The three-year WMAP(WMAP3), combined  with other cosmological
observations from galaxy clustering and Type Ia Supernova (SNIa),
prefers a non-vanishing running of the primordial spectral index
independent of the low CMB multipoles. Motivated by this feature we
study cosmological constraint on the neutrino mass, which severely
depends on what prior we adopt for the spectral shape of primordial
fluctuations,
 taking possible  running into account.
As a result we find a more stringent constraint on the sum of the
three neutrino masses, $m_\nu < 0.76$eV (2 $\sigma$), compared with
$m_\nu < 0.90$eV (2 $\sigma$) for the case where power-law prior is
adopted to the primordial spectral shape.

\end{abstract}

\maketitle

\hskip 1.6cm PACS number(s): 98.80.Cq \vskip 0.4cm

The three year Wilkinson Microwave Anisotropy Probe observations
(WMAP3) \cite{Spergel:2006hy,Page:2006hz,Hinshaw:2006,Jarosik:2006,WMAP3IE}
have marked another milestone on the precision cosmology of
the Cosmic Microwave
Background (CMB) Radiation.  The simplest six-parameter power-law
$\Lambda$CDM cosmology is in remarkable agreement with WMAP3
together with the large scale structure(LSS) of galaxy clustering
as measured by 2dF \cite{Cole:2005sx} and SDSS \cite{Tegmark:2003uf}
and with the Type Ia Supernova (SNIa) as measured by the Riess
"gold" sample \cite{Riess:2004nr} and the first year SNLS \cite{snls}.
This agreement between the above ``canonical'' cosmological model
and observations can be used to test a number of possible new physics,
such as the equation of state of dark energy, neutrino masses, time
variation of fundamental constants, etc.

Among them, the constraint on the neutrino or hot dark matter mass
can be obtained from the free-streaming modification of the
transfer function of the matter power spectrum.
We should note, however, that if one allowed any shape
of primordial spectrum, the free-streaming effect could easily be
compensated by some nontrivial shape of the primordial spectrum,
so that one cannot obtain sensible limit on the neutrino mass.
That is,
we can obtain a nontrivial bound on neutrino mass if and only
if we adopt some prior on the shape of primordial power spectrum
such as a simple power-law.  From the above argument, we
expect that as we allow more degrees
of freedom on the primordial spectrum beyond a power-law, the
constraint on the neutrino mass would be less stringent in general.

As for the shape of the primordial power spectrum,
it is noteworthy that a significant deviation has been observed
by WMAP3 from the simplest
Harrison-Zel'dovich spectrum, and that this feature is more
eminent with the combination of all the currently available CMB, LSS
and SNIa (dubbed the case of "All" in  \cite{WMAP3IE}). Moreover, a
nontrivial negative running of the scalar spectral index $\alpha_s$,
whose existence was studied even before WMAP epoch
 \cite{Lewis:2002ah,Feng:2003nt}, was
favored by the first-year
WMAP papers  \cite{wmap1a,wmap1b,wmap1c}.
But its preference was somehow diminished
as corrections to the likelihood
functions were made  \cite{Slosar:2004fr}.
However, the new
WMAP3 data prefers again a negative running
in the "All" combination  \cite{WMAP3IE}.

If  confirmed, a nonvanishing  running of $\alpha_s$ would
not only constrain inflationary cosmology significantly
 \cite{Feng:2003mk,Kawasaki:2003zv,Yamaguchi:2003fp,Yamaguchi:2004tn,Chen:2004nx,
Ballesteros:2005eg}, but also affect the cosmological constraint
on the neutrino mass. In the LSS power spectrum, the effect of
massive neutrino may be compensated by a nonvanishing
 running of primordial spectrum. On the other hand, if
negativeness of the running is established, it will lead to even
more stringent constraints on neutrino mass compared with fittings
in the constant scalar spectral index($n_s$) cosmology, because
 both of them lead to a damped power on small scales.

The actual problem, however, cannot be solved by the above simple
one-to-one correspondence, because the effects of a nonvanishing
neutrino mass on CMB is much less dramatic than a nonvanishing
 running $\alpha_s$.
Alternatively,
in the scales probed by CMB,
especially near the third peak, there is a large degeneracy between
$\alpha_s$ and the matter density $\Omega_m$
 \cite{Dodelson:1995es,Hannestad:2006zg,Lesgourgues:2006nd,Ichikawa:2004zi,Fukugita:2006rm}.

Hence in the concordance analysis the correlation between running
and neutrino mass needs to be addressed in a combined study of CMB,
LSS and SNIa, where SNIa helps significantly to determine the matter
density.  We report the results of such a combined analysis in this
paper using Markov Chain Monte Carlo method in constraining
the total neutrino mass, $m_\nu = \sum_{i=1}^3 m_{\nu i}
=94.4\Omega_\nu h^2$eV.
Here $\Omega_\nu$ is the density parameter of the
neutrino and $h$ is the Hubble constant in unit of 100km/s/Mpc.

To break possible degeneracy among the cosmological
parameters, we make a global fit to the forementioned current data
with the publicly available Markov Chain Monte Carlo
package \texttt{cosmomc} \cite{Lewis:2002ah,IEMCMC}. Our most general
parameter space is:
\begin{equation}\label{para}
    \textbf{p}\equiv(\omega_{b}, \omega_{c}, \Theta_S, \tau, m_{\nu}, n_{s}, \alpha_{s},  \log[10^{10} A_{s}])
\end{equation}
where $\omega_{b}=\Omega_{b}h^{2}$ and
$\omega_{c}=\Omega_{c}h^{2}$ are the physical baryon and cold dark
matter densities relative to critical density, $\Theta_S$
characterizes the ratio of the sound horizon and angular diameter
distance, $\tau$ is the optical depth and $A_{s}$ is defined as
the amplitude of initial power spectrum. The pivot scale for
$n_{s}$  and $\alpha_{s}$ is chosen at $k=0.05$ Mpc$^{-1}$.

Assuming a flat Universe and in terms of the Bayesian
analysis, we vary the above 8 parameters and fit the theory to the
observational data with the MCMC method. For CMB we have only
adopted the WMAP3.  The
bias factors of LSS have been used as nuisance parameters and hence
essentially we have used only the shapes of 2dF and SDSS power
spectra.
As for the SNIa data, while the WMAP team uses both SNLS
 and Riess
sample simultaneously in their ``All'' dataset, here we adopt only
one of them, namely, the Riess "gold" sample rather than combining
with SNLS. This is because these two groups use somewhat different
methods in their analysis and it would not be appropriate to put
them together simply\cite{footnt}.

Regarding the first-year WMAP data Bridle et al.\  \cite{Bridle:2003sa}
found that the claimed preference of a negative
$\alpha_s$ was merely due to the lowest WMAP multipoles.
In order to probe the sensitivity of the running to the lower
multipoles we analyze
 the running and neutrino properties using the CMB data
with and without the contributions of lower  multipoles
which suffer from large cosmic variance.
Specifically, we truncate naturally at $l=24$ given the current
likelihood of
WMAP3 \cite{Spergel:2006hy,Page:2006hz,Hinshaw:2006,WMAP3IE}.

As a result we find a more stringent constraint on the neutrino mass
in the presence of running
 compared with the analysis with constant $n_s$.
We also
find that currently the preference of a negative running is
fairly
independent of the WMAP3 low CMB quadrupoles and hence relatively
robust.

\begin{table*}
TABLE 1. Mean with 1$\sigma$ (2$\sigma$) constrains on the spectral
index, the running, and the neutrino mass based on
 LSS and SNIa with WMAP3(with/without
$l<24$ CMB contributions) and with/without introducing a running of
the primordial spectral index $\alpha_{s}$ and with/without massive
neutrinos. The last column shows the corresponding reduction of
$\chi^2$ values compared with the power law $\Lambda$CDM cosmology.

\begin{center}
\begin{tabular}{|c|ccc|ccc|}

  \hline
&\multicolumn{3}{c|}{Normal WMAP3}&\multicolumn{3}{c|}{$l<24$ dropped}\\

&$\nu$&$\alpha_s$&$\nu$+$\alpha_s$ &$\nu$&$\alpha_s$&$\nu$+$\alpha_s$ \\

\hline

$n_s$ &$0.947^{+0.016}_{-0.016}(^{+0.033}_{-0.032})$
&$0.880^{+0.038}_{-0.037}(^{+0.074}_{-0.073})$
&$0.899^{+0.040}_{-0.041}(^{+0.077}_{-0.078})$
&$0.942^{+0.021}_{-0.022}(^{+0.051}_{-0.039})$
&$0.862^{+0.049}_{-0.049}(^{+0.096}_{-0.095})$
&$0.899^{+0.058}_{-0.059}(^{+0.110}_{-0.113})$ \\

$\alpha_s$     & set to $0$
&$-0.051^{+0.025}_{-0.025}(^{+0.051}_{-0.048})$
&$-0.037^{+0.028}_{-0.028}(^{+0.053}_{-0.053})$
& set to $0$
&$-0.072^{+0.043}_{-0.042}(^{+0.083}_{-0.089})$
&$-0.045^{+0.052}_{-0.050}(^{+0.098}_{-0.101})$ \\

$m_{\nu}$ &$0.460^{+0.113}_{-0.460}(^{+0.437}_{-0.460})$
& set to $0$
&$0.314^{+0.079}_{-0.314}(^{+0.442}_{-0.314})$
&$0.470^{+0.131}_{-0.470}(^{+0.499}_{-0.470})$
& set to $0$
&$0.378^{+0.099}_{-0.378}(^{+0.523}_{-0.378})$  \\

\hline
$\Delta \chi^2 $ & -2 & $-3.4$ & $-4.1$ & -1 & $-2.1$ & $-2.2$ \\

\hline

\end{tabular}
\end{center}
\end{table*}

In Table 1 we show the mean 1$\sigma$ (2$\sigma$) constrains on the
relevant cosmological parameters combining 2dF, SDSS and Riess
"gold" sample with WMAP3. We have addressed the cases with/without
$l<24$ CMB contributions, with/without introducing  $\alpha_{s}$ and
with/without massive neutrinos. The last column shows the
corresponding reduction of $\chi^2$ values compared with the power
law $\Lambda$CDM cosmology. For normal  LSS $+$ SNIa $+$ WMAP in the
7 parameter fittings we get $\alpha_{s} =
-0.0512^{+0.0506}_{-0.0480}$ at 2$\sigma$. Our results are
considerably less stringent than the "All" combination by WMAP team
\cite{WMAP3IE}, as we have not included SNLS and other CMB
observations. On the other hand the preference of non-vanishing
$\alpha_{s}$ is larger than 2$\sigma$ in both cases. The preference
to negative running still remains even if we drop the WMAP3 $l<24$
contributions, when we get $\alpha_{s} =
-0.0717^{+0.0833}_{-0.0886}$ at 2$\sigma$. This may imply a negative
running is indeed preferred nontrivially  in the combination of
WMAP+LSS+SNIa, even without the presence of WMAP3 small $l$
contributions. This can also be understood through the Akaike
information criterion (AIC \cite{AIC}), which is defined as defined
as
\begin{equation}
AIC=-2\ln{\mathcal{L}}+2k,
\end{equation}
where $\mathcal{L}$ is the maximum likelihood, $k$ is the number of
parameters of the model and $N$ is the number of datapoints. From
Table I we can find the reduction of $\Delta \chi^2 \equiv -2\Delta
\ln{\mathcal{L}}$ is larger than 2 in the case of running compared
with the simple power law $\Lambda$CDM cosmology, hence we have
smaller values of AIC with running, although currently not strong to
decisive levels.

\begin{figure}[htbp]
\begin{center}
\includegraphics[scale=0.55]{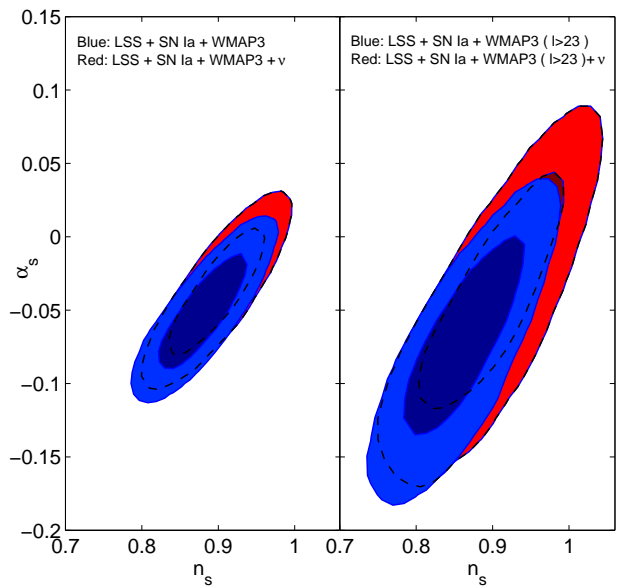}
\caption{ Two dimensional posterior constraints on the primordial
spectral index versus its running at $k=0.05$ Mpc$^{-1}$, using 2dF,
SDSS, SNIa and WMAP3(with/without $l<24$ CMB contributions) and
with/without the presence of massive neutrinos. \label{fig:fig1}}
\end{center}
\end{figure}

The preference of a negative running can also be obtained in the
two-dimensional posterior contours of $n_s-\alpha_{s}$, as depicted
in Fig.~\ref{fig:fig1}. For the 7-parameter case with one additional
parameter of $\alpha_{s}$, although the contour without WMAP3 low
$l$ contributions is larger than that in the left panel, a constant
$n_s$ lies close to the 2$\sigma$ lines in both cases. The enlarged
contour without $l<24$ is easily understood due to the $n_s-\tau$
degeneracy.

The correlation between $m_{\nu}$ and the shape of the primordial
spectrum is obvious, as can be seen from Table 1 and Fig.~
\ref{fig:fig1}. In the presence of massive neutrinos the error
bars on $n_s$ and $\alpha_{s}$ get increased disregard with or
without small $l$ CMB contributions. And it is noteworthy in the
presence of massive neutrinos, a scale-invariant primordial
spectrum is consistent with the observations at $\sim 2 \sigma$ if
we drop the small $l$ WMAP3 contributions.

\begin{figure}[htbp]
\begin{center}
\includegraphics[scale=0.8]{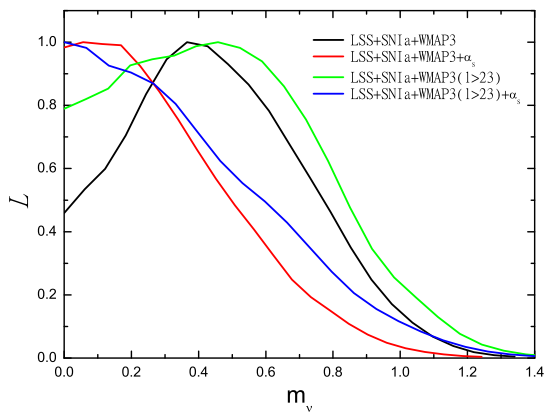}
\caption{ One dimensional posterior constraints on the sum of
neutrino masses, using 2dF, SDSS, SNIa and WMAP3(with/without
$l<24$ CMB contributions) and with/without introducing a running
of the primordial spectral index $\alpha_{s}$. Note due to the
non-gaussian distribution of $m_{\nu}$, the LSS $+$ SNIa $+$ WMAP3
(black line) seems to indicate a nonzero neutrino mass, but it is
not the case, as can be seen in Table 1. \label{fig:fig2}}
\end{center}
\end{figure}

We find that, while almost all of the remaining parameters get less
stringently constrained, the neutrino mass is an exception: a more
tightened bound on $m_{\nu}$ is achieved in the presence of a
nonzero $\alpha_{s}$ than the case with constant $n_s$, as shown in
Table 1 and in the one dimensional constraints as in
Fig.~\ref{fig:fig2}.
This has shown the fact that running is indeed
strongly correlated with neutrino mass, which is mainly due to the
physics on LSS.
\if
During the epoch of structure formation the
gravitational instability of the fluctuations plays a crucial role.
Massive neutrinos serve to prevent the instabilities on distances
smaller than the neutrinos' free streaming scale. On small scales
the growth of the fluctuations is suppressed by
 \cite{Lesgourgues:2006nd}
\begin{equation}
\left( {\Delta P\over P }\right) \approx
-8{\Omega_\nu\over\Omega_m}\approx-0.8 \left( { m_\nu \over 1 e V}
\right) \left( {0.1 N_{\nu} \over\Omega_m h^2}\right)\,.
\label{eq:detP}
\end{equation}
And the physics of running on the LSS power spectrum  is
straightforward on the linear scales: \be P(k) = P_s(k) T^2(k)
,\label{mpkns} \ee where running is present in the sector of the
primordial spectrum $P_s(k)$ and massive neutrinos essentially
contribute to the matter transfer function $T(k)$.
\fi

\begin{figure}[htbp]
\begin{center}
\includegraphics[scale=0.55]{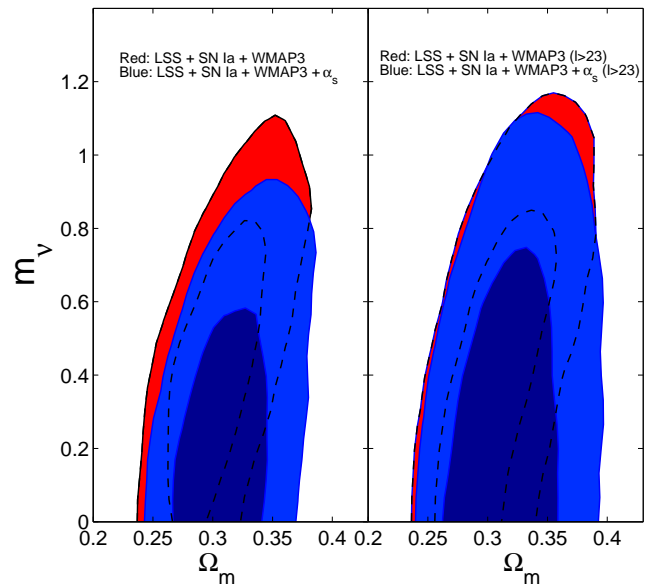}
\caption{ Two dimensional posterior constraints on the sum of
neutrino masses versus matter density, using 2dF, SDSS, SNIa and
WMAP3(with/without $l<24$ CMB contributions) and with/without
introducing a running of the primordial spectral index
$\alpha_{s}$.
 \label{fig:fig3}}
\end{center}
\end{figure}

In Fig.~\ref{fig:fig3} we display the two dimensional posterior
constraints on the sum of neutrino masses versus matter
density in the same case as Fig.~\ref{fig:fig2}. While the allowed
parameter space for $\Omega_m$ is significantly enlarged  in the presence
of running, neutrino mass is constrained stringently in cases with
nonzero $\alpha_{s}$. The correlation between $\Omega_m$ and
$m_{\nu}$ is rather strong and the accumulation of the observational
data, such as SNAP, PLANCK and SDSS will help significantly to break
the degeneracy, to detect the features of the primordial spectrum as
well as the nature of neutrinos.

It has been claimed that a running of the spectral index will be
excluded in the presence of SDSS Lyman $\alpha$ observations
 \cite{Seljak:2004sj,Hannestad06x,Viel:2006yh,Seljak:2006bg}, the systematics of Lyman $\alpha$
data are relatively less constrained
\cite{Spergel:2006hy,Lesgourgues:2006nd} and we leave this in a
separate investigation to be reported in  \cite{Feng:2006ui}, and
detailed analysis with other additional possible degeneracies in
 \cite{toappear2}.

For many years neutrinos have played a fundamental role in both
physics and astrophysics,
and provided venues of new physics such
as  the parity violation and oscillations with tiny masses.
Surely
neutrinos will continue to play a crucial role for our understanding
of the Universe. While the resulting reduction on neutrino mass in
this paper is less than 0.2 eV, such an effect would hopefully help
to change our understandings on the ultimate detection of neutrino
mass with future cosmological surveys and the difference is already
larger than the low limit of neutrino mass from oscillation
experiments. On the other hand there are also some mild tensions in
the determinations of the background cosmological parameters with
current CMB, LSS and SNIa. We may still need some better
understandings on each data-set before entering the precision
cosmology  \cite{Bridle:2003yz} and in cases when all of the
observations have similar tendencies in the preference of a negative
running one can hopefully get more eminent effects in the probe of
neutrino mass. 

The distinctive feature probed in the current paper will also open
new windows on relevant studies, such as probing neutrino mass with
a nonzero running in gravitational lensing surveys, N-body
simulations in the presence of running and massive neutrinos.

{\bf{Acknowledgments:}} 
We acknowledge the use of the
Legacy Archive for Microwave Background Data Analysis (LAMBDA).
Support for LAMBDA is provided by the NASA Office of Space Science.
We have performed our numerical analysis on the Shanghai
Supercomputer Center(SSC). We used a modified version of
CAMB \cite{Lewis:1999bs,IEcamb} which is based on
CMBFAST \cite{cmbfast,IEcmbfast}. We thank Steen Hannestad, Antony
Lewis, Chris Lidman, Hiranya Peiris and Pengjie Zhang for helpful
discussions. The work of J.Y. is supported partially by the JSPS
Grant-in-Aid for Scientific Research No. 16340076 and B.F. is by the
JSPS fellowship program. This work is supported in part by National
Natural Science Foundation of China under Grant Nos. 90303004,
10533010 and 19925523 and by Ministry of Science and Technology of
China under Grant No. NKBRSF G19990754.

{}


\begin{thebibliography}{}
\bibitem{footnt}
Currently the SNLS and Riess "gold" sample are comparable in the
determination of cosmological parameters, as shown in Ref.
\cite{WMAP3IE}, and each dataset has its own nice features. In the
present work we use only the Riess sample.
\end{thebibliography}
\end{document}